\documentstyle[12pt,openbib]{article}
\newcommand{\de}{\delta}
\newcommand{\p}{\prime}
\newcommand{\la}{\lambda}
\newcommand{\La}{\Lambda}

\newcommand{\be}{\beta}

\newcommand{\s}{\Sigma}
\newcommand{\ca}{\Xi}
\newcommand{\g}{\gamma}

\newcommand{\f}{\frac}

\begin{document}
\title{\bf Uses of QCD sumrules (QCDSR) from  different points of view.}
\vskip .5cm
\author{Jishnu Dey $^{1,3}$, Mira Dey  $^{2,3,4}$, 
 \\ and \\Mahendra Sinha Roy$^{4}$}
\vspace{.5 cm}
\date{\today }
\maketitle
{\it Abstract} :
We support the idea that the baryon, B with mass $M_B$,  couples to its
current with a coupling $\la _{B}^2 \;\sim \; 0.71 \; M_B^6$ from an
analysis of magnetic moment sum rules. And we find a sum rule among the
experimental magnetic moments which is independent of the parameters of QCDSR.
\vskip .5cm

(1) Azad Physics Centre, Dept. of Physics, Maulana Azad College, Calcutta 700
013, India\\ 
(2) ICTP, Trieste, Italy 341000
(3) {\it{ Work supported in part by DST grant no. SP/S2/K18/96, Govt.
of India, \\ permanent address : 1/10 Prince Golam Md. Road,
Calcutta 700 026, India, email : deyjm@giascl01.vsnl.net.in}}.
(4) Dept. of Physics, Presidency College, Calcutta 700 073,
India \\
\newpage
It was suggested recently \cite {dey}, that the coupling of the octet and the
decuplet baryons to its interpolating current in the framework of the SVZ
QCDSR (see \cite{dey} for earlier references) scales with the baryons masses
with a constant $C = 1.37$, as follows :
\begin{equation}
\la _{B}^2 \;\sim \; C \; M_B^6
\label{eq:1}
\end{equation}
In our notation $\lambda _B$ is defined through the matrix element
\begin{equation}
\langle 0 |{\cal J}_B| B \rangle = \la _B \; v^{(r)}
\label{eq:2}
\end{equation}
where ${\cal J}_B$ is the interpolating baryon current and $v^{(r)}$ is the
usual Dirac spinor for polarization $r$, normalized to the baryon mass $M_B$
as :
\begin{equation}
{\bar v^{(r)}}v^{(r)} = 2 M_B.
\label{eq:3}
\end{equation}  
We call eq.(\ref {eq:1}) the scaling law, (SL in short). Very general arguments as well as a specific model was put forward in support of this SL in \cite{dey}. The SL is able to connect QCDSR for various quantities (for example mass, magnetic moments, transitions or differences thereof) of different baryons. These connections, which are the common sumrules, to differentiate from QCDSR, are to be checked with experimental numbers. In the present paper this task is achieved for the differences in magnetic moments of three multiplets : $(p-n)$, $(\s^+ - \s^-)$ and $(\ca^- - \ca^0)$, an experimentally verifiable sumrule is established among these.
 
This SL was checked in \cite{lee}, for decuplet baryons with a
constant $C = 0.71$, in eq.(\ref{eq:1}) and the $\s^*$, $\ca^*$ and $\Omega$
couplings fit very well. The isobar $\Delta$ is off the curve and Lee gets an
overestimate for the $\Delta(1232)$ mass as well, so that it is not clear
which mass to use in the abscissa of the curve given by eq.(\ref{eq:1}). He
observed that  ``Given the great importance of a scaling law between baryon
current couplings and their masses and its phenomenological consequences,
more investigations are clearly needed to resolve the deviations."

The results of \cite{dey} were based on QCDSR for masses which are now
modified partially. In \cite{lein}, uncertainties in the predictions
of QCDSR calculations have been analysed comprehensively in
$\rho$-meson and nucleon two-point functions.  Indeed, the results are
pretty disturbing for QCDSR practitioners, as Ref.\cite{lein}
indicates that reducing these uncertainties requires both a better
determination of the vacuum condensates and a better determination of
the sum rules themselves. The nucleon coupling deduced in \cite{lein}  using
Monte-Carlo -based analysis still deviates from the SL \cite{lee}. To
investigate the scaling law further, we look into the QCDSR for magnetic moment
differences.   

Thus our outlook is quite different from that of \cite{dey,lee,lein} in so far
as we do not look at mass QCDSR, but at QCDSR for magnetic moment differences
among octet -baryon isomultiplets to find support for the scaling argument of
the coupling . We propose a magnetic moment sum rule which is experimentally
verifiable and does not involve QCDSR parameters.
\newpage
Our starting point is the observation that the magnetic moments, for the baryons $B$, 
can be written in the form \cite{cpw}  
\begin{equation}
\mu _B = constant \; \;(1 + \de _B) \;\f{e\hbar}{2cM_B}
\label{eq:4}
\end{equation}

where the $constant$ is 8/3 for proton $p$, $\s ^+$ and  -4/3 for neutron 
$n$, $\s ^-$, $\ca ^-$ and $\ca ^0$ \cite{cpw}. The $\de _B $ are small
numbers and for the differences $(\de_p-\de_n)$, $(\de_{\s ^{+}} - \de_ {\s
^{-}})$ and $(\de_{\ca ^{-}} - \de_{\ca ^{0}})$ the QCD sumrules simplify
very much \cite{cwps}. The susceptibility terms cancel out. For completeness
we write, as an example, the QCDSR for the nucleon (i.e. proton - neutron)
from \cite{cwps}.  $$\be _{N}^2 [\de _p - \de _n + (A_p -
A_n)M^2].exp(-M_N^2/M^2) =$$
\begin{equation}
(7/2-2)\f{M^2 b}{192 L^{4/9}} + (-1/2 -1) \f{M^2 b}{288 L^{\f{4}{9}}} \xi 
+ (-1/2- 4) a^2 L^{\f{4}{9}}/72
\label{eq:5}
\end{equation}
with 
$$\la _B ^2 = 8 \be _B ^2 $$
$$a = - 2\pi ^2 <\bar q q>$$
$$b = <g_s^2 G^2>$$ 
$$L = \f{ln(\f{M^2}{\La_{QCD}^2})}{ln(\f{\mu ^2}{\La_{QCD}^2})}$$
$$\xi = ln(\f{M^2}{\La^2})-1-\g$$
where $\La$ the cut-off, $\mu$ the renormalization point, $\La_{QCD}$  and
$\g$, the Euler-Mascheroni constant are 0.5 GeV, 0.5 GeV, 0.1 GeV and 0.577
respectively.  Similar equations are given in \cite{cwps} for $\s$ and $\ca$.

We simplify the problem further by writing down the equations for $\de _B -
\de _{B^\p}$ after operating with $(1 - M^2.\f{\partial}{\partial M^2})$,
where M is the unknown of the QCDSR, the Borel parameter. The last operation,
first used by Ioffe and Smilga, \cite{is}, gets rid of the intermediate state
contributions marked $A_B, A_{B^\p}$ in \cite{is, cwps}. Abbreviating by
\begin{equation}
DL = \f{4}{9}ln(\f{\mu ^2}{\La _{QCD}^2})
\label{eq:6}
\end{equation}
and
\begin{equation}
E = exp(\f{M_B^2}{M^2})
\label{eq:7}
\end{equation}
the equations are  :
\begin{equation}
f1(M^2)=q{\f{M^2}{288}}(\xi \f{M_B^2}{M^2} E L^{-\f{4}{9}}-{\f{E}{{L}^
{\f{4}{9}}}}+\xi L^{-\f{13}{9}})
\label{eq:8}
\end{equation}

\begin{equation}
f2(M^2)=[\f{M_B^2}{M^2} E .L ^{-\f{4}{9}}+DL.E .L ^{-\f{13}{9}}]\f{M^2}{8}.s
\label{eq:9}
\end{equation}
\begin{equation}
f3(M^2)=[\f{M_B^2}{M^2} E .L ^{-\f{4}{9}}+DL.E .L ^{-\f{13}{9}}].M^2.\f{p}
{192}
\label{eq:10}
\end{equation}

\begin{equation}
f4(M^2)=L ^{\f{4}{9}}.(E +\f{M_B^2}{M^2} E -DL.{\f{E }{L }}).\f{r}{72}
\label{eq:11}
\end{equation}
\begin{equation}
 (\de _B - \de _{B^\p}) = [ f1(M^2).b +  f2(M^2) m_s.a
+ f3(M^2).b + f4(M^2).a^2 ]/\be _{B}^2
\label{eq:12}
\end{equation}
The coefficients $p,q,r,s$ are given in Table 1. 

Thus the difference of magnetic moments are simple functions which {\it only
depend on the condensate values} $a$ and $b$ and the contribution of various
higher order terms and the intermediate excitations are easily gotten rid of.
Of course there still are too many parameters since the various couplings
$\be _B$ appears in the RHS of each equation. We suggest that one should use
the SL to explore if the $\be _B$ can be expressed in terms of one
another. And thus with three experimental mass differences one can determine
the coefficient of the cubic SL connecting the $\be _B$, and $a$ and
$b$.

We reemphasize that in this paper we find that the difference of magnetic
moments is an ideal ground for searching for the truth of SL rather
than mass sum rules which are riddled with many problems.

The next point we wish to stress is that over the years the value of the
condensates $a$ and $b$ have changed a lot and now carry large uncertainties.
Thus \cite{lein,lee} quote the values $a = (0.52 \pm 0.05)\; GeV^3$ and $b =
\;(1.2 \pm 0.6) \; GeV^4$. In the following, we will attempt to find some
suitable values of these parameters from magnetic moment differences.

RHS of eq. (\ref{eq:12}) is a function of the Borel parameter, M. This is
shown in Fig. 1 and the RHS matches the experimental numbers only at $M \;
=\; M_B$. We allow the $\be _B$ in Table 2 to have a small scatter from eq.
(\ref{eq:1}) with $ C\;=\;0.71$,  and find that with $a \sim 0.47 \; GeV^3$
and $b \sim 1.7 \; GeV^4$ we can fit the three magnetic moment differences.
Next we proceed to find unique solutions of $b$ and $a$. 

Ideally one should be able to work with any value of the Borel parameter M in QCDSR. But from Fig. 1 we see that this is not possible, the QCDSR curves for the differences in $\de _B$ will cross the experimental straight line only at one point. We accept the situation as such and only demand that the value of $\de _B - \de _{B^\p}$ fits experimental numbers (Table 2) at $M = M_B$. We observe from eq.  (\ref{eq:9}) and Table 1 that $m_s a$ occurs only in  $\ca$ : 
$$0.42215\; b - 4.8837\; a^2 \;= \;(\de_p-\de_n)$$ 
$$-0.13416\;b + 1.294 a^2  = \; (\de_{\s ^{+}} - \de_{\s ^{-}}) $$.
\begin{equation}
0.235827\; b - 1.613178\; a^2 - \; 8.228 (m_s a) = (\de_{\ca ^{-}} - \de_{\ca
^{0}})
\label{eq:main}
\end{equation}

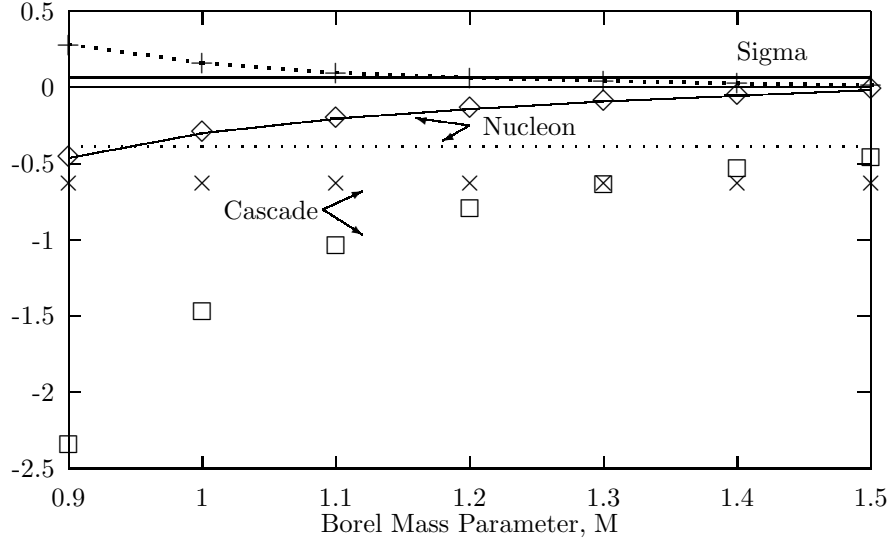
\begin{figure}
\begin{center}
\setlength{\unitlength}{0.240900pt}
\ifx\plotpoint\undefined\newsavebox{\plotpoint}\fi
\begin{picture}(1200,500)(0,0)
\font\gnuplot=cmr10 at 10pt
\gnuplot
\sbox{\plotpoint}{\rule[-0.200pt]{0.400pt}{0.400pt}}%
\put(176.0,712.0){\rule[-0.200pt]{303.534pt}{0.400pt}}
\put(176.0,113.0){\rule[-0.200pt]{4.818pt}{0.400pt}}
\put(154,113){\makebox(0,0)[r]{-2.5}}
\put(1416.0,113.0){\rule[-0.200pt]{4.818pt}{0.400pt}}
\put(176.0,233.0){\rule[-0.200pt]{4.818pt}{0.400pt}}
\put(154,233){\makebox(0,0)[r]{-2}}
\put(1416.0,233.0){\rule[-0.200pt]{4.818pt}{0.400pt}}
\put(176.0,353.0){\rule[-0.200pt]{4.818pt}{0.400pt}}
\put(154,353){\makebox(0,0)[r]{-1.5}}
\put(1416.0,353.0){\rule[-0.200pt]{4.818pt}{0.400pt}}
\put(176.0,472.0){\rule[-0.200pt]{4.818pt}{0.400pt}}
\put(154,472){\makebox(0,0)[r]{-1}}
\put(1416.0,472.0){\rule[-0.200pt]{4.818pt}{0.400pt}}
\put(176.0,592.0){\rule[-0.200pt]{4.818pt}{0.400pt}}
\put(154,592){\makebox(0,0)[r]{-0.5}}
\put(1416.0,592.0){\rule[-0.200pt]{4.818pt}{0.400pt}}
\put(176.0,712.0){\rule[-0.200pt]{4.818pt}{0.400pt}}
\put(154,712){\makebox(0,0)[r]{0}}
\put(1416.0,712.0){\rule[-0.200pt]{4.818pt}{0.400pt}}
\put(176.0,832.0){\rule[-0.200pt]{4.818pt}{0.400pt}}
\put(154,832){\makebox(0,0)[r]{0.5}}
\put(1416.0,832.0){\rule[-0.200pt]{4.818pt}{0.400pt}}
\put(176.0,113.0){\rule[-0.200pt]{0.400pt}{4.818pt}}
\put(176,68){\makebox(0,0){0.9}}
\put(176.0,812.0){\rule[-0.200pt]{0.400pt}{4.818pt}}
\put(386.0,113.0){\rule[-0.200pt]{0.400pt}{4.818pt}}
\put(386,68){\makebox(0,0){1}}
\put(386.0,812.0){\rule[-0.200pt]{0.400pt}{4.818pt}}
\put(596.0,113.0){\rule[-0.200pt]{0.400pt}{4.818pt}}
\put(596,68){\makebox(0,0){1.1}}
\put(596.0,812.0){\rule[-0.200pt]{0.400pt}{4.818pt}}
\put(806.0,113.0){\rule[-0.200pt]{0.400pt}{4.818pt}}
\put(806,68){\makebox(0,0){1.2}}
\put(806.0,812.0){\rule[-0.200pt]{0.400pt}{4.818pt}}
\put(1016.0,113.0){\rule[-0.200pt]{0.400pt}{4.818pt}}
\put(1016,68){\makebox(0,0){1.3}}
\put(1016.0,812.0){\rule[-0.200pt]{0.400pt}{4.818pt}}
\put(1226.0,113.0){\rule[-0.200pt]{0.400pt}{4.818pt}}
\put(1226,68){\makebox(0,0){1.4}}
\put(1226.0,812.0){\rule[-0.200pt]{0.400pt}{4.818pt}}
\put(1436.0,113.0){\rule[-0.200pt]{0.400pt}{4.818pt}}
\put(1436,68){\makebox(0,0){1.5}}
\put(1436.0,812.0){\rule[-0.200pt]{0.400pt}{4.818pt}}
\put(176.0,113.0){\rule[-0.200pt]{303.534pt}{0.400pt}}
\put(1436.0,113.0){\rule[-0.200pt]{0.400pt}{173.207pt}}
\put(176.0,832.0){\rule[-0.200pt]{303.534pt}{0.400pt}}
\put(806,23){\makebox(0,0){Borel Mass Parameter, M}}
\put(827,652){\makebox(0,0)[l]{Nucleon}}
\put(1226,765){\makebox(0,0)[l]{Sigma}}
\put(420,520){\makebox(0,0)[l]{Cascade}}
\put(176.0,113.0){\rule[-0.200pt]{0.400pt}{173.207pt}}
\multiput(802.68,650.92)(-0.879,-0.496){45}{\rule{0.800pt}{0.120pt}}
\multiput(804.34,651.17)(-40.340,-24.000){2}{\rule{0.400pt}{0.400pt}}
\put(764,628){\vector(-2,-1){0}}
\multiput(793.96,652.58)(-3.598,0.492){21}{\rule{2.900pt}{0.119pt}}
\multiput(799.98,651.17)(-77.981,12.000){2}{\rule{1.450pt}{0.400pt}}
\put(722,664){\vector(-4,1){0}}
\multiput(575.00,520.58)(1.092,0.497){55}{\rule{0.969pt}{0.120pt}}
\multiput(575.00,519.17)(60.989,29.000){2}{\rule{0.484pt}{0.400pt}}
\put(638,549){\vector(2,1){0}}
\multiput(575.00,518.92)(0.789,-0.498){77}{\rule{0.730pt}{0.120pt}}
\multiput(575.00,519.17)(61.485,-40.000){2}{\rule{0.365pt}{0.400pt}}
\put(638,480){\vector(3,-2){0}}
\put(176,601){\usebox{\plotpoint}}
\multiput(176.00,601.58)(2.711,0.498){75}{\rule{2.254pt}{0.120pt}}
\multiput(176.00,600.17)(205.322,39.000){2}{\rule{1.127pt}{0.400pt}}
\multiput(386.00,640.58)(4.629,0.496){43}{\rule{3.752pt}{0.120pt}}
\multiput(386.00,639.17)(202.212,23.000){2}{\rule{1.876pt}{0.400pt}}
\multiput(596.00,663.58)(7.167,0.494){27}{\rule{5.700pt}{0.119pt}}
\multiput(596.00,662.17)(198.169,15.000){2}{\rule{2.850pt}{0.400pt}}
\multiput(806.00,678.58)(9.026,0.492){21}{\rule{7.100pt}{0.119pt}}
\multiput(806.00,677.17)(195.264,12.000){2}{\rule{3.550pt}{0.400pt}}
\multiput(1016.00,690.59)(12.203,0.489){15}{\rule{9.433pt}{0.118pt}}
\multiput(1016.00,689.17)(190.421,9.000){2}{\rule{4.717pt}{0.400pt}}
\multiput(1226.00,699.59)(12.203,0.489){15}{\rule{9.433pt}{0.118pt}}
\multiput(1226.00,698.17)(190.421,9.000){2}{\rule{4.717pt}{0.400pt}}
\put(176,601){\raisebox{-.8pt}{\makebox(0,0){$\Diamond$}}}
\put(386,640){\raisebox{-.8pt}{\makebox(0,0){$\Diamond$}}}
\put(596,663){\raisebox{-.8pt}{\makebox(0,0){$\Diamond$}}}
\put(806,678){\raisebox{-.8pt}{\makebox(0,0){$\Diamond$}}}
\put(1016,690){\raisebox{-.8pt}{\makebox(0,0){$\Diamond$}}}
\put(1226,699){\raisebox{-.8pt}{\makebox(0,0){$\Diamond$}}}
\put(1436,708){\raisebox{-.8pt}{\makebox(0,0){$\Diamond$}}}
\put(176,619){\usebox{\plotpoint}}
\multiput(176,619)(20.756,0.000){11}{\usebox{\plotpoint}}
\multiput(386,619)(20.756,0.000){10}{\usebox{\plotpoint}}
\multiput(596,619)(20.756,0.000){10}{\usebox{\plotpoint}}
\multiput(806,619)(20.756,0.000){10}{\usebox{\plotpoint}}
\multiput(1016,619)(20.756,0.000){10}{\usebox{\plotpoint}}
\multiput(1226,619)(20.756,0.000){10}{\usebox{\plotpoint}}
\put(1436,619){\usebox{\plotpoint}}
\sbox{\plotpoint}{\rule[-0.400pt]{0.800pt}{0.800pt}}%
\put(176,728){\usebox{\plotpoint}}
\put(176.0,728.0){\rule[-0.400pt]{303.534pt}{0.800pt}}
\sbox{\plotpoint}{\rule[-0.500pt]{1.000pt}{1.000pt}}%
\put(176,779){\usebox{\plotpoint}}
\multiput(176,779)(20.560,-2.839){11}{\usebox{\plotpoint}}
\multiput(386,750)(20.703,-1.479){10}{\usebox{\plotpoint}}
\multiput(596,735)(20.740,-0.790){10}{\usebox{\plotpoint}}
\multiput(806,727)(20.750,-0.494){10}{\usebox{\plotpoint}}
\multiput(1016,722)(20.752,-0.395){10}{\usebox{\plotpoint}}
\multiput(1226,718)(20.753,-0.296){10}{\usebox{\plotpoint}}
\put(1436,715){\usebox{\plotpoint}}
\put(176,779){\makebox(0,0){$+$}}
\put(386,750){\makebox(0,0){$+$}}
\put(596,735){\makebox(0,0){$+$}}
\put(806,727){\makebox(0,0){$+$}}
\put(1016,722){\makebox(0,0){$+$}}
\put(1226,718){\makebox(0,0){$+$}}
\put(1436,715){\makebox(0,0){$+$}}
\sbox{\plotpoint}{\rule[-0.600pt]{1.200pt}{1.200pt}}%
\put(176,149){\raisebox{-.8pt}{\makebox(0,0){$\Box$}}}
\put(386,358){\raisebox{-.8pt}{\makebox(0,0){$\Box$}}}
\put(596,461){\raisebox{-.8pt}{\makebox(0,0){$\Box$}}}
\put(806,520){\raisebox{-.8pt}{\makebox(0,0){$\Box$}}}
\put(1016,557){\raisebox{-.8pt}{\makebox(0,0){$\Box$}}}
\put(1226,582){\raisebox{-.8pt}{\makebox(0,0){$\Box$}}}
\put(1436,599){\raisebox{-.8pt}{\makebox(0,0){$\Box$}}}
\sbox{\plotpoint}{\rule[-0.500pt]{1.000pt}{1.000pt}}%
\put(176,561){\makebox(0,0){$\times$}}
\put(386,561){\makebox(0,0){$\times$}}
\put(596,561){\makebox(0,0){$\times$}}
\put(806,561){\makebox(0,0){$\times$}}
\put(1016,561){\makebox(0,0){$\times$}}
\put(1226,561){\makebox(0,0){$\times$}}
\put(1436,561){\makebox(0,0){$\times$}}
\end{picture}
\caption{ Experimental $(\de _B - \de _{B^\p})$ are shown as a solid line,
dotted line and crosses, the topmost for $B \;=\;\s$, the middle one for $B
\;=\;N$ and the lowest one for $B \;=\;\Xi$. The QCDSR results, as a function
of Borel parameter M, crosses at the respective $M \;=\;M_B$.}
\end{center}
\end{figure}

\begin{figure}
\begin{center}
\setlength{\unitlength}{0.240900pt}
\ifx\plotpoint\undefined\newsavebox{\plotpoint}\fi
\sbox{\plotpoint}{\rule[-0.200pt]{0.400pt}{0.400pt}}%
\begin{picture}(1200,500)(0,0)
\font\gnuplot=cmr10 at 10pt
\gnuplot
\sbox{\plotpoint}{\rule[-0.200pt]{0.400pt}{0.400pt}}%
\put(220.0,113.0){\rule[-0.200pt]{292.934pt}{0.400pt}}
\put(220.0,113.0){\rule[-0.200pt]{4.818pt}{0.400pt}}
\put(198,113){\makebox(0,0)[r]{0}}
\put(1416.0,113.0){\rule[-0.200pt]{4.818pt}{0.400pt}}
\put(220.0,203.0){\rule[-0.200pt]{4.818pt}{0.400pt}}
\put(198,203){\makebox(0,0)[r]{2}}
\put(1416.0,203.0){\rule[-0.200pt]{4.818pt}{0.400pt}}
\put(220.0,293.0){\rule[-0.200pt]{4.818pt}{0.400pt}}
\put(198,293){\makebox(0,0)[r]{4}}
\put(1416.0,293.0){\rule[-0.200pt]{4.818pt}{0.400pt}}
\put(220.0,383.0){\rule[-0.200pt]{4.818pt}{0.400pt}}
\put(198,383){\makebox(0,0)[r]{6}}
\put(1416.0,383.0){\rule[-0.200pt]{4.818pt}{0.400pt}}
\put(220.0,473.0){\rule[-0.200pt]{4.818pt}{0.400pt}}
\put(198,473){\makebox(0,0)[r]{8}}
\put(1416.0,473.0){\rule[-0.200pt]{4.818pt}{0.400pt}}
\put(220.0,562.0){\rule[-0.200pt]{4.818pt}{0.400pt}}
\put(198,562){\makebox(0,0)[r]{10}}
\put(1416.0,562.0){\rule[-0.200pt]{4.818pt}{0.400pt}}
\put(220.0,652.0){\rule[-0.200pt]{4.818pt}{0.400pt}}
\put(198,652){\makebox(0,0)[r]{12}}
\put(1416.0,652.0){\rule[-0.200pt]{4.818pt}{0.400pt}}
\put(220.0,742.0){\rule[-0.200pt]{4.818pt}{0.400pt}}
\put(198,742){\makebox(0,0)[r]{14}}
\put(1416.0,742.0){\rule[-0.200pt]{4.818pt}{0.400pt}}
\put(220.0,832.0){\rule[-0.200pt]{4.818pt}{0.400pt}}
\put(198,832){\makebox(0,0)[r]{16}}
\put(1416.0,832.0){\rule[-0.200pt]{4.818pt}{0.400pt}}
\put(220.0,113.0){\rule[-0.200pt]{0.400pt}{4.818pt}}
\put(220,68){\makebox(0,0){0.9}}
\put(220.0,812.0){\rule[-0.200pt]{0.400pt}{4.818pt}}
\put(372.0,113.0){\rule[-0.200pt]{0.400pt}{4.818pt}}
\put(372,68){\makebox(0,0){1}}
\put(372.0,812.0){\rule[-0.200pt]{0.400pt}{4.818pt}}
\put(524.0,113.0){\rule[-0.200pt]{0.400pt}{4.818pt}}
\put(524,68){\makebox(0,0){1.1}}
\put(524.0,812.0){\rule[-0.200pt]{0.400pt}{4.818pt}}
\put(676.0,113.0){\rule[-0.200pt]{0.400pt}{4.818pt}}
\put(676,68){\makebox(0,0){1.2}}
\put(676.0,812.0){\rule[-0.200pt]{0.400pt}{4.818pt}}
\put(828.0,113.0){\rule[-0.200pt]{0.400pt}{4.818pt}}
\put(828,68){\makebox(0,0){1.3}}
\put(828.0,812.0){\rule[-0.200pt]{0.400pt}{4.818pt}}
\put(980.0,113.0){\rule[-0.200pt]{0.400pt}{4.818pt}}
\put(980,68){\makebox(0,0){1.4}}
\put(980.0,812.0){\rule[-0.200pt]{0.400pt}{4.818pt}}
\put(1132.0,113.0){\rule[-0.200pt]{0.400pt}{4.818pt}}
\put(1132,68){\makebox(0,0){1.5}}
\put(1132.0,812.0){\rule[-0.200pt]{0.400pt}{4.818pt}}
\put(1284.0,113.0){\rule[-0.200pt]{0.400pt}{4.818pt}}
\put(1284,68){\makebox(0,0){1.6}}
\put(1284.0,812.0){\rule[-0.200pt]{0.400pt}{4.818pt}}
\put(1436.0,113.0){\rule[-0.200pt]{0.400pt}{4.818pt}}
\put(1436,68){\makebox(0,0){1.7}}
\put(1436.0,812.0){\rule[-0.200pt]{0.400pt}{4.818pt}}
\put(220.0,113.0){\rule[-0.200pt]{292.934pt}{0.400pt}}
\put(1436.0,113.0){\rule[-0.200pt]{0.400pt}{173.207pt}}
\put(220.0,832.0){\rule[-0.200pt]{292.934pt}{0.400pt}}
\put(10,472){\makebox(0,0){Coupling Square }}
\put(828,23){\makebox(0,0){Baryon Mass }}
\put(220.0,113.0){\rule[-0.200pt]{0.400pt}{173.207pt}}
\put(281,141){\raisebox{-.8pt}{\makebox(0,0){$\Diamond$}}}
\put(659,222){\raisebox{-.8pt}{\makebox(0,0){$\Diamond$}}}
\put(725,244){\raisebox{-.8pt}{\makebox(0,0){$\Diamond$}}}
\put(851,249){\raisebox{-.8pt}{\makebox(0,0){$\Diamond$}}}
\put(956,337){\raisebox{-.8pt}{\makebox(0,0){$\Diamond$}}}
\put(1182,483){\raisebox{-.8pt}{\makebox(0,0){$\Diamond$}}}
\put(1393,749){\raisebox{-.8pt}{\makebox(0,0){$\Diamond$}}}
\put(281,135){\usebox{\plotpoint}}
\multiput(281,135)(20.428,3.675){19}{\usebox{\plotpoint}}
\multiput(659,203)(19.690,6.563){3}{\usebox{\plotpoint}}
\multiput(725,225)(19.132,8.048){7}{\usebox{\plotpoint}}
\multiput(851,278)(18.095,10.167){6}{\usebox{\plotpoint}}
\multiput(956,337)(15.887,13.356){14}{\usebox{\plotpoint}}
\multiput(1182,527)(12.406,16.640){17}{\usebox{\plotpoint}}
\put(1393,810){\usebox{\plotpoint}}
\end{picture}
\caption{The coupling square $\la _{B}^2$ as a function of the baryon mass.
The dashed curve is the equation eq.(1) with C = 0.71.}
\end{center}
\end{figure}
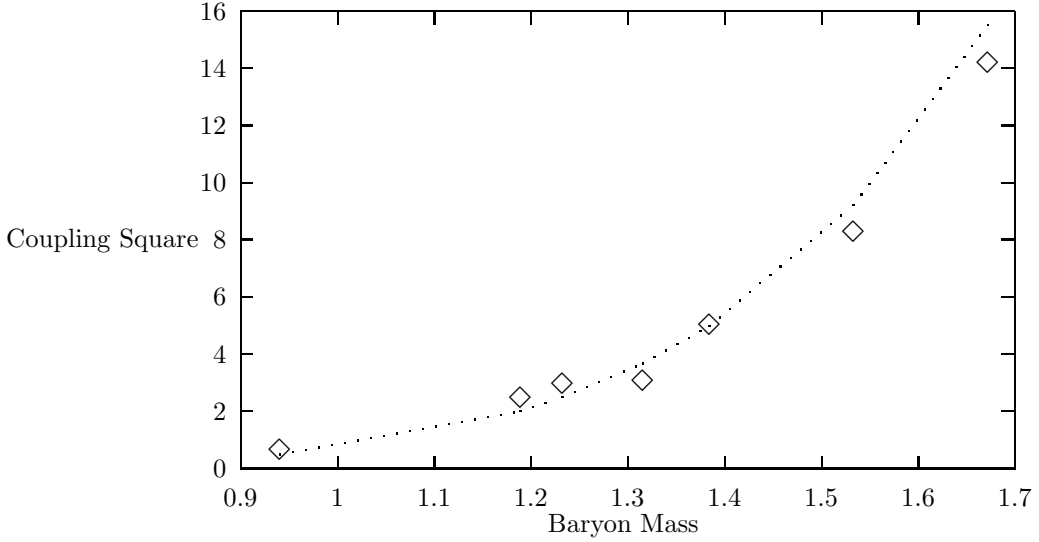

On solving first two of these equations one gets a set of values of $ b$ and
$ a^2$, which by our choice of the couplings gives consistent results :
\begin{equation}
 b \;= \;1.695 GeV^4
\label{eq:14}
\end{equation} 
\begin{equation}
 a \;= \; 0.475 GeV^3.
\label{eq:15}
\end{equation}
These values of $a$ and $b$ and the equation (\ref{eq:main}) combined give us
the strange quark mass $m_s\; =\; 169.8\; MeV $ which is reasonable.

Thus we conclude that to get results for magnetic moments, consistent with
experiment, at least for $M = M_B$, one must choose the parameters $a$ and
$b$ which are close to the extreme limits of the error range given by
\cite{lein,lee}.  

We now proceed to investigate the SL for decuplet baryons with the values of $a$ and $b$ given above. We find in Table 3 that in particular the isobar $\Delta(1232)$ has a much significantly smaller $\la_{\Delta}$ given by :
\begin{equation}
\la ^2 _\Delta =  [\f{4}{3}a E_1({\f {M^2}{M_{\Delta}^2}}) L^{\f{16}{27}}M^4 - \f{2}{3}a E_0({\f {M^2}{M_{\Delta}^2}})m_0^2 L^{\f{2}{27}}M^2 - {\f{b}{18}} a L^{\f{16}{27}}] e^{\f {M^2}{M_{\Delta}^2}}/M_{\Delta}
\label{eq:16}
\end{equation}
where $E_n(x) = 1- e^{-x} \s_n x^n/n!$, and $m_0^2 \;= \;0.72\; GeV^2$.
Using the experimental mass, $\la_{\Delta}$ fits into the SL very well in Fig. 2. Reversing the argument one can state that if one were to trust the SL we advocate, the isobar mass problem may be in better shape. Note that Lee \cite{lee} uses a coupling square in his tables which must be multiplied by a factor of 2 to compare with ours.

Now comes our central result : a sumrule which is independent of the
steps of QCDSR through which it was derived. 

We have shown that to get results for magnetic moments, (or at least the difference between the moments of two baryons in the same multiplet, consistent with experiment at $M \;=\; M_B$, one must choose the parameters $a$ and $b$ which are close to the extreme limits of the error range given by \cite{lein,lee}. But then, we have these parameters given in terms of the differences $(\de_p - \de_n)$, $(\de_{\s ^{+}} - \de_{\s ^{-}})$ and
$(\de_{\ca ^{-}} - \de_{\ca ^{0}})$. We can solve equations (\ref{eq:main}) either for $a$ or $b$ to get a consistent sumrule : 
\begin{equation}
(\de_p-\de_n)\; = \;-0.8153 - 1.2273 \; (\de_{\ca ^{-}} - \de_{\ca ^{0}}) -
5.3045 (\de_{\s ^{+}} - \de_{\s ^{-}}).
\label{eq:17}
\end{equation} 

Note that magnetic moment experiments are being refined all the time and that
$\s$ and $\ca$ magnetic moments were revised recently and are different from
the ones quoted in \cite{cpw, cwps}. It will be interesting to see if the
above sumrule will stand the test of time.

In summary we have used the scaling of the baryon coupling to its current
to predict a sumrule among difference of octet baryon magnetic moments. We
also indicate how the SL can be used to ameliorate other difficulties of
QCDSR technique.

We believe that the SLs can be extended to charmed and beauty baryons. It may
be possible to establish connections between decays from baryons and check
them with experiments.

The authors are grateful for several e-mails from Drs. Lee and Leinweber. One
of the authors (MD) wishes to thank the UNESCO, IAEA and ICTP for hospitality
at Trieste, Italy. 

\newpage
\begin{table}
\begin{center}
\caption {Coefficients p, q, r and s}
\vskip 1cm
\begin{tabular}{c|c|c|c|c}
\hline
System & p &  q & r & s  \\
\hline
p - n & 1.5  & - 1.5 & - 4.5 &  0 \\
$\s ^+ - \s^-$ & -1.5  & 1.5 & 4.5 &  0 \\
$\ca ^- - \ca ^0 $ & 3  & -3 & -9 -18 f &  -6 \\
\hline
\end{tabular}
\end{center}
\end{table}

\begin{table}
\begin{center}
\caption {Experimental numbers $\de _B -\de _{B^\p}$  for octet baryons along
with our squared couplings $\be _B^2$}
\vskip 1cm
\begin{tabular}{c|c|c|c|c}
\hline
$M_B \,(GeV)$ &$\be _B^2$ & System & $\de _B -
\de _{B^\p}$   \\
\hline
0.940 & 0.0766 &p - n & - 0.388  \\
1.189 & 0.3037 &$\s^ + - \s ^- $ & 0.065  \\
1.315 & 0.3776 &$\ca ^- -\ca ^0$  & - 0.629  \\
\hline
\end{tabular}
\end{center}
\end{table}

\begin{table}
\begin{center}
\caption {Decuplet couplings }
\vskip 1cm
\begin{tabular}{c|c|c|c|c}
\hline
$M_B \,(GeV)$ & continuum w $(GeV)$ & Our $\la _B $ & 
$\la _B$ from \cite{lee} \\
\hline
1.232 & 1.65 & 1.458 & 2.26 $\pm$ 0.89  \\
1.384 & 1.80 & 2.492 & 2.83 $\pm$ 0.32  \\
1.533 & 2.0  & 4.122 & 4.32 $\pm$ 0.47  \\
1.672 & 2.3 & 7.081 & 7.19 $\pm$ 0.75  \\
\hline
\end{tabular}
\end{center}
\end{table}

\begin{thebibliography}{}
\bibitem{dey} J. Dey, M. Dey, T. Frederico and L. Tomio, Mod. Phys. Lett. A
12 (1997) 2193.
\bibitem{lee} F. X. Lee, Phys. Rev. D57 (1997) 322.
\bibitem{lein} D. B. Leinweber, Ann. Phys. 254 (1997) 328.
\bibitem{lein2} D. B. Leinweber, The truth about nucleon sum rules, manuscript
in preparation.
\bibitem{cpw} C. B. Chiu, J. Pasupathy and S. J. Wilson, Phys. Rev. D 33
(1986) 1961. 
\bibitem{cwps} C. B. Chiu, S. J. Wilson, J. Pasupathy and J. P. Singh, Phys.
Rev. D 36 (1987) 1553.
\bibitem{is} B. L. Ioffe and A. V. Smilga, Nucl. Phys. B 232 (1984) 109. 
\end{thebibliography}
\end{document}